\documentclass[fleqn,usenatbib]{mnras}

\usepackage{newtxtext,newtxmath}

\usepackage[T1]{fontenc}

\DeclareRobustCommand{\VAN}[3]{#2}
\let\VANthebibliography\thebibliography
\def\thebibliography{\DeclareRobustCommand{\VAN}[3]{##3}\VANthebibliography}

\usepackage{graphicx}	
\usepackage{amsmath}	
\defcitealias{Prole2022}{LP22}

\title[Primordial magnetic fields in Population III star formation: a magnetised resolution study]{Primordial magnetic fields in Population III star formation: a magnetised resolution study}

\author[L. R. Prole]{
Lewis R. Prole,$^{1}$\thanks{E-mail: Prolel@cardiff.ac.uk}
Paul C. Clark,$^{1}$
Ralf S. Klessen,$^{2,3}$
Simon C. O. Glover,$^{2}$
R\"{u}diger Pakmor$^{4}$
\\
$^{1}$Cardiff University School of Physics and Astronomy\\
$^{2}$Zentrum f{\"u}r Astronomie, Universit{\"a}t Heidelberg, Institut f{\"u}r Theoretische Astrophysik, Albert-Ueberle-Str. 2, 69120 Heidelberg, Germany\\
$^{3}$Interdisziplin{\"a}res Zentrum fur Wissenschaftliches Rechnen, INF 205, D-69120, Heidelberg, Germany\\
$^{4}$ Max-Planck-Institut f\"{u}r Astrophysik, Karl-Schwarzschild-Str. 1, D-85741 Garching, Germany
}

\date{Accepted XXX. Received YYY; in original form ZZZ}

\pubyear{2015}

\begin{document}
\label{firstpage}
\pagerange{\pageref{firstpage}--\pageref{lastpage}}
\maketitle

\begin{abstract}

Population III stars form in groups due to the fragmentation of primordial gas. While uniform magnetic fields have been shown to support against fragmentation in present day star formation, it is unclear whether realistic $k^{3/2}$ primordial fields can have the same effect. We bypass the issues associated with simulating the turbulent dynamo by introducing a saturated magnetic field at equipartition with the velocity field when the central densities reaches $10^{-13}$ g cm$^{-3}$. We test a range of sink particle creation densities from 10$^{-10}$-10$^{-8}$ g cm$^{-3}$. Within the range tested, the fields did not suppress fragmentation of the gas and hence could not prevent the degree of fragmentation from increasing with increased resolution. The number of sink particles formed and total mass in sink particles was unaffected by the magnetic field across all seed fields and resolutions. The magnetic pressure remained sub-dominant to the gas pressure except in the highest density regions of the simulation box, where it became equal to but never exceeded gas pressure. Our results suggest that the inclusion of magnetic fields in numerical simulations of Pop III star formation is largely unimportant.
\end{abstract}




\section{Introduction}
Initial investigations into Population III (Pop III) star formation found that the first stars were massive (>100M$_\odot$: \citealt{Bromm1999}) and formed in isolation (e.g. \citealt{Haiman1996}). However, simulations by \cite{Clark2011}, which included a primordial heating and cooling network, proved that Pop III clouds were susceptible to fragmentation in the presence of sub-sonic turbulence, resulting in groups of stars with lower masses and a broadened IMF. This fragmentation behaviour has been seen in many studies since (e.g. \citealt{Smith2011, Greif2012, Stacy2013, Stacy2014, Susa2019, Wollenberg2019}). In \cite{Prole2022}, we showed that the degree of fragmentation in primordial gas increases as the maximum density of the simulation is increased, likely only converging once the gas enters its adiabatic regime at $\sim 10^{-4}$ g cm$^{-3}$. Crucially, these studies did not include the magnetic fields predicted to exist in the early Universe (e.g.  \citealt{Schober2015}).

Numerical investigations including magnetic fields in a present-day star formation setting show that the effects of magnetic tension and pressure can drastically change the outcome of the collapse (e.g. \citealt{Machida2005,Machida2006,Price2007,Machida2008,Hennebelle2008,Hennebelle2008b, Hennebelle2011,Burzle2011b}). The large-scale galactic magnetic field is uniform over the scales involved in star formation, so these studies employ uniform magnetic fields either aligned or misaligned with the axis of rotation. For ideal magneto-hydrodynamics (ideal MHD: i.e. for a fluid which is a perfect conductor with zero resistivity) magnetic field lines are frozen into the fluid and can be advected and distorted as the fluid moves \citep{Alfven1942}. This \emph{flux-freezing} allows conservation of magnetic flux during a gravitational collapse, leading to a natural amplification of the magnetic field strength within the cloud as $\propto \rho^{2/3}$. After the formation of a disc, rotation can distort an initially poloidal magnetic field into an increasingly toroidal field with successive disc rotations (see Figure 1 from \cite{Machida2008a}).  From \emph{Ferraro's law of isorotation} \citep{Ferraro1937}, when the field lines are distorted due to rotation of the disc, magnetic tension acts to correct the distortion by transferring angular momentum to the slower regions, known as \emph{magnetic braking}. Simulations show that this removal of angular momentum from the disc can act to prevent fragmentation and delay the onset of star formation (e.g. \citealt{Hennebelle2008b, Hennebelle2011,Burzle2011b}), while \cite{Price2007} instead attribute reduced fragmentation to the isotropic magnetic pressure forces rather than magnetic tension. Angular momentum can also be removed from the system via outflows or jets. Jets are believed to be launched via a magneto-centrifugal mechanism when magnetic field lines make an angle of $<60$ degrees with the disc. This can happen when the initially uniform and poloidal magnetic fields twist with the disc. Centrifugal forces drive the plasma out of the accretion disc along the field lines and the magnetic stress associated with the twisting forces cause the outflow to become collimated into a jet (e.g.\citealt{Blandford1982,Lynden-Bell1996}). Simulations typically produce a low velocity flow known as a molecular outflow from from first adiabatic core (e.g. \citealt{Machida2005,Hennebelle2008}), followed by a high velocity flow known as an optical jet from the second stellar core (e.g. \citealt{Machida2006,Machida2008}).

Population III star formation simulations have also been performed in the presence of uniform magnetic fields. Three-dimensional MHD nested grid simulations by \cite{Machida2008a} showed that fragmentation and jet behaviour depended on the ratio of the initial rotational to magnetic energy. Magnetically-dominated scenarios resulted in jets without fragmentation while rotationally-dominated models resulted in fragmentation without jets. \cite{Turk2012} studied the amplification of initially uniform magnetic fields within Pop III star-forming regions by performing cosmological simulations, finding a refinement criteria of 64 zones per Jeans length was needed to capture dynamo action. They also found that better-resolved simulations possess higher infall velocities, increased temperatures inside 1000 AU, decreased molecular hydrogen content in the innermost region and suppressed disc formation. \cite{Machida2013} performed resistive MHD simulations with an initially uniform magnetic field, finding that for initial field strengths above $10^{-12}$ G, angular momentum around the protostar is transferred by both magnetic braking and protostellar jets. In this case, the gas falls directly on to the protostar without forming a disc, forming a single massive star. Recently, \cite{Sadanari2021} performed a similar study with uniform magnetic fields, including all the relevant cooling processes and non-equilibrium chemical reactions up to protostellar densities, finding that magnetic effects become important for field strengths greater than $10^{-8}$ G. While these studies show that uniform magnetic fields can change the outcome of Pop III star formation, the magnetic fields present during the formation of Pop III stars likely were highly disordered, unlike the large-scale fields present in the galactic ISM today.

Multiple mechanisms of generating magnetic fields in the early Universe are covered in literature. Generation of magnetic fields is predicted by Faraday’s law if an electric field caused by electron pressure gradients has a curl, which is known as the Biermann battery \citep{Biermann1950}. Alternatively, an anisotropic distribution of electron velocities in a homogeneous, collisionless plasma supports unstable growth of electromagnetic waves, known as the Weibel instability \citep{Weibel1959}. Magnetic fields may have also been produced by cosmological density fluctuations \citep{Ichiki2006} or by cosmic rays from Pop III supernovae which induce currents, the curl of which are capable of generating magnetic fields \citep{Miniati2011}. Irrespective of their origin, magnetic seed fields are believed to have eventually grown into the large scale galactic magnetic fields observed today (e.g. \citealt{Kulsrud1990}). It is suggested that the amplification begins during the collapse of the gas that formed the first stars (e.g. \citealt{Kulsrud1997,Schleicher2010, Silk2006}), which form within low mass dark matter (DM) halos \citep{Couchman1986}.  Along with the natural $\propto \rho^{2/3}$ amplification from flux-freezing, the small-scale turbulent dynamo \citep{Vainshtein1980} converts the kinetic energy of an electrically conductive medium into magnetic energy, amplifying the magnetic field when turbulence is generated during the collapse of the DM minihalo. Magnetic fields amplified by turbulent motions produce $P(k) \propto k^{3/2}$ power spectra \citep{Kazantsev1968}, where the energy increases for smaller spatial scales, in contrast to the standard turbulent velocity power spectrum which has more energy on larger spatial scales. The turbulent dynamo ends once the field has saturated at an equilibrium between kinetic and magnetic energy. This is expected to happen on timescales shorter than the collapse time of the halo (e.g. \citealt{Schober2012}). While these fields are small scale and chaotic, they can still assist in resisting fragmentation due to the isotropic nature of the magnetic pressure, which contributes to a quasi-isotropic acoustic-type wave along with the gas pressure.

The magnetic field strength resulting from the small-scale turbulent dynamo depends on the efficiency of the dynamo process, which depends on the Reynolds number \citep{Sur2010} and hence, in simulations, the resolution used \citep{Haugen2004}. This  results in underestimated dynamo amplification of the magnetic fields in numerical simulations. Despite underestimated amplification, \cite{Schleicher2010} showed that magnetic fields are significantly enhanced before the formation of a protostellar disc, where they can change the fragmentation properties of the gas and the accretion rate. \cite{Federrath2011} ran MHD dynamo amplification simulations, testing the effects of varying the Jeans refinement criterion from 8-128 cells per Jeans length on the field amplification during collapse. They ran their simulations up to density $\sim 10^{-13}$ g cm$^{-3}$, finding that dynamo amplification was only seen when using a refinement criteria of 32 cells per Jeans length and above. The resulting power spectrum was of the $k^{3/2}$ Kazantsev type. The field was still amplified for smaller refinement criteria as $\rho^{2/3}$ i.e. due to flux freezing during the gravitational collapse. They found that the scale where magnetic energy peaks shifts to smaller scales as resolution was increased. The velocity power spectra indicated that gravity-driven turbulence exhibits an effective driving scale close to the Jeans scale. For the 128 cells per Jeans length run, the magnetic field strength reached $\sim 1$mG when the central region of collapse reached 10$^{-13}$ g cm$^{-3}$. \cite{Schober2012a} found that the dynamo saturates at $\sim 10^{-24}$ g cm$^{-3}$ at strengths of $10^{-6}$G, afterwhich the strength continues to increase due to flux freezing only. \cite{Schober2015} presented a scale-dependent saturation model, where the peak scale of the magnetic spectrum moves to larger spatial scales as the field amplifies, until reaching a peak scale at saturation. The peak energy scale at saturation depends on the driving scale of the turbulence, which in turn depends on the slope of the velocity spectrum. The ratio of magnetic to kinetic energy $\text{E}_{\text{B}} / \text{E}_{\text{k}}$ at dynamo saturation varies between studies; in \cite{Schober2015}  $\text{E}_{\text{B}} /  \text{E}_{\text{k}}=0.0134$ for pure Burgers compressible turbulence ($P(k) \propto k^{-2}$), increasing to 0.304 as the turbulence was mixed more with incompressible modes into pure incompressible Kolmogorov turbulence ($P(k) \propto k^{-5/3}$).  \cite{Federrath2011a} also investigated the saturation level of the turbulent dynamo, finding values in the range 0.001 to 0.6, while \cite{Haugen2004} found a value of 0.4. The theoretical upper limit is a ratio of 1 i.e. equipartition between the velocity and magnetic field.

While most Pop III MHD simulations have implemented unrealistic uniform magnetic fields, recently \cite{Sharda2020} performed MHD simulations with a Kazantsev $k^{3/2}$ spectrum, finding suppressed fragmentation in the presence of strong magnetic fields, resulting in a reduction in the number of first stars with masses low enough that they might be expected to survive to the present-day. However, they only ran the simulations up to a maximum density of $\sim 10^{-11}$ g cm$^{-3}$ and spatial resolution of 7.6 au. If primordial gas becomes stable to fragmentation at densities of $10^{-4}$ g cm$^{-3}$ at T$\sim 10^{4}$ K corresponding to a Jeans scale of $0.024$ au, the study can not sufficiently resolve the degree of fragmentation primordial gas experiences. Furthermore, the limited resolution means that the smallest scale magnetic fields are artificially uniform on scales roughly 300 times larger than the stellar cores. Also, collision-induced emission kicks in to become the dominant cooling process at $\sim 10^{-10}$ g cm$^{-3}$ and dissociation of H$_{2}$ molecules provides effective cooling after $\sim 10^{-8}$ g cm$^{-3}$ \citep{Omukai2005}, so the study has not included a  full chemical treatment. There are further concerns that an under-resolved initial magnetic field would lead to collapse within a region that does not properly sample the $k^{3/2}$ power spectrum. In the worst case scenario the collapse could occur within a region of uniform field, significantly increasing the ability of the field to suppress fragmentation.

This investigation attempts to bypass the effects of underestimating dynamo growth and the complications associated with resolution by introducing a high resolution saturated field after an initial non-magnetized phase of collapse. We zoom-in on the most central region for the magnetised phase and set the magnetic field strength to the theoretical maximum by choosing the dynamo saturation energy to be at equipartition with the velocity field i.e.  E$_{\text{mag}}$/E$_{\text{kinetic}} =1$. In \cite{Prole2022} (hereafter \citetalias{Prole2022}), we concluded that in the purely hydrodynamic case, the degree of disc fragmentation would not converge until the gas becomes adiabatic at $10^{-4}$ g cm$^{-3}$, shifting the IMF to smaller mass stars as the maximum density increases. This paper aims to test if realistic primordial magnetic fields can provide the necessary support against fragmentation to converge the sink particle mass spectrum before the formation of the adiabatic core. Due to the increased computing resources required for MHD simulations, the maximum density of the resolution test was chosen to catch all of the relevant chemical processes, the last of which is the dissociation of H$_{\text{2}}$ molecules at $\sim 10^{-8}$ g cm$^{-3}$. The smallest cells in these simulations resolve spatial scales of 0.128 au, roughly 5 times larger than the stellar core, improving on previous studies by 3 orders of magnitude in density and by a factor of 60 in spatial resolution.

The structure of the paper is as follows. In Section \ref{sec:nummeth} we discuss the numerical method, including the simulation code {\sc Arepo}, the chemical network, implementation of ideal MHD and use of sink particles. In Section \ref{sec:sims} we give our initial conditions and explain the two-stage zoom in simulations. In Section \ref{sec:frag} we discuss the fragmentation behaviour of the primordial gas under the influence of magnetic fields, while in Section \ref{sec:field} we discuss the magnetic field behaviour. In Section \ref{sec:comparison} we compare our findings with previous studies before discussing caveats in Section \ref{sec:caveats} and concluding in Section \ref{sec:conclusion}.

\section{Numerical method }
\label{sec:nummeth}

\subsection{{\sc Arepo}}
We have performed two-stage zoom-in MHD simulations with the moving mesh code {\sc Arepo} \citep{Springel2010} with a primordial chemistry set-up. {\sc Arepo} combines the advantages of AMR and smoothed particle hydrodynamics (SPH: \citealt{Monaghan1992}) with a mesh made up of a moving, unstructured, Voronoi tessellation of discrete points. {\sc Arepo} solves hyperbolic conservation laws of ideal hydrodynamics with a finite volume approach, based on a second-order unsplit Godunov scheme. Several different Riemann solvers are available, including an exact solver and the HLLD solver \citep{Miyoshi2005} used in this work.

\subsection{Chemistry}
Our initial conditions begin once the gas has already been cooled by molecular hydrogen to $\sim$200K at $\sim$10$^{-20}$ g cm$^{-3}$. An overview of the chemical reactions and cooling processes relevant for the further collapse of the gas is given in \cite{Omukai2005}. Briefly, once the collapse reaches $\sim$10$^{-16}$ g cm$^{-3}$, three-body H$_{2}$ formation begins to convert most of the atomic hydrogen into molecular hydrogen, accompanied by a release of energy that heats the gas to $\sim$1000K. At 10$^{-14}$ g cm$^{-3}$, the gas becomes optically thick to H$_2$ cooling. At $\sim$10$^{-10}$ g cm$^{-3}$, collision induced cooling kicks in and the gas becomes optically thick to continuum at $\sim$10$^{-8}$ g cm$^{-3}$. From there, H$_2$ dissociation provides cooling until all of the H$_{2}$ is depleted and the gas becomes adiabatic at $\sim$10$^{-4}$ g cm$^{-3}$.

We use the same chemistry and cooling as \cite{Wollenberg2019}, which is based on the fully time-dependent chemical network described in the appendix of \cite{Clark2011}, but with updated rate coefficients, as summarised in \cite{Schauer2017}. The network has 45 chemical reactions to model primordial gas made up of 12 species: H, H$^{+}$, H$^{-}$, H$^{+}_{2}$ , H$_{2}$, He, He$^{+}$, He$^{++}$, D, D$^{+}$, HD and free electrons.  Included in the network are: H$_{2}$ cooling (including an approximate treatment of the effects of opacity), collisionally-induced H$_{2}$ emission, HD cooling, ionisation and recombination, heating and cooling from changes in the chemical make-up of the gas and from shocks, compression and expansion of the gas, three-body H$_{2}$ formation and heating from accretion luminosity. The network switches off tracking of deuterium chemistry at densities above 10$^{-16}$~g~cm$^{-3}$, instead assuming that the ratio of HD to H$_{2}$ at these densities is given by the cosmological D to H ratio of 2.6$\times$10$^{-5}$. The adiabatic index of the gas is computed as a function of chemical composition and temperature.

\subsection{MHD in {\sc Arepo}}
\label{sec:MHDarepo}
Ideal MHD is implemented in {\sc Arepo} by converting the ideal MHD equations into a system of conservation laws as 
\begin{equation}
\frac{\delta \vec{U}}{\delta t} + \bigtriangledown \cdot (\vec{F}) = 0
\end{equation}
where 

\begin{equation}
U=\begin{pmatrix}
\rho \\
\rho \vec{v} \\
\rho e \\
\vec{B} \\
\end{pmatrix}	        \; \; \;
F(U)=\begin{pmatrix}
\rho \vec{v} \\
\rho \vec{v} \vec{v}^{T}+p-\vec{B}\vec{B}^T \\
\rho e\vec{v}+p\vec{v}-\vec{B}(\vec{v}\cdot \vec{B}) \\
\vec{B} \vec{v}^T - \vec{v} \vec{B}^T\\
\end{pmatrix}
\end{equation}

where $\rho$, $\vec{v}$ and $\vec{B}$ are the local gas density, velocity and magnetic field strength, $p =p_{\text{gas}} + \frac{1}{2}B^2$ is the total gas pressure, $e =u + \frac{1}{2}\vec{v}^2 + \frac{1}{2\rho}B^2$  is the total energy, where $u$ is the thermal energy per unit mass. These conservation laws reduce to ideal hydrodynamics when $B=0$. The equations are solved with a second-order accurate finite-volume scheme \citep{Pakmor2011}. 

Electric charges have no magnetic analogues, so the net outflow of the magnetic field through any arbitrary closed surface is zero. Magnetic fields are therefore solenoidal vector fields, i.e. 

\begin{equation}
\bigtriangledown \cdot \vec{B} = 0
\end{equation}

It is common for $\bigtriangledown \cdot \vec{B}$ errors to arise in numerical simulations. This may result in non-physical results, such as plasma transport orthogonal to the magnetic field lines. Constrained transport \citep{Evans1988} methods have been developed to restrict $\bigtriangledown \cdot \vec{B}$ to 0, and a version of this algorithm has been implemented in {\sc Arepo} by \cite{Mocz2016}. However, the current implementation of constrained transport in {\sc Arepo} requires that all of the mesh cells share the same global timestep, i.e.\ it does not permit the use of hierarchical time-stepping. This dramatically reduces the computational efficiency of the code for high dynamical range problems, such as the one considered here, and renders it impractical to use this technique for our current simulations. Instead, to deal with $\bigtriangledown \cdot \vec{B}$ errors we use hyperbolic divergence cleaning for the MHD equations whereby the divergence constraint is coupled with the conservation laws by introducing a generalized Lagrange multiplier \citep{Powell1995,Dedner2002}.

\subsection{Sink particles}
A given structure of gas can support itself thermally from gravitational collapse at scales up to the Jeans length $\lambda_{\text{J}}$, which is a function of gas density and temperature. If the Jeans length of the gas within a cell of the {\sc Arepo} mesh is allowed to become smaller than the cell size, it cannot support itself and artificial numerical effects occur. Numerical simulations must refine the mesh as the gas gets denser to ensure that the local Jeans length is resolved, but cannot do so indefinitely. When the simulation reaches a threshold density, a point mass known as a sink particle is introduced to represent the gas. We use the same sink particle treatment as \cite{Wollenberg2019} and \cite{Tress2020}. A cell is converted into a sink particle if it satisfies three criteria: 1) it reaches a threshold density; 2) it is sufficiently far away from pre-existing sink particles so that their accretion radii do not overlap; 3) the gas occupying the region inside the sink is gravitationally bound and collapsing. Likewise, for the sink particle to accrete mass from surrounding cells it must meet two criteria: 1) the cell lies within the accretion radius; 2) it is gravitationally bound to the sink particle. A sink particle can accrete up to 90$\%$ of a cell's mass, above which the cell is removed and the total cell mass is transferred to the sink.

The sink particle treatment also includes the accretion luminosity feedback from \cite{Smith2011}. The internal luminosity of the star is not included in this work, which does not affect the results as our simulations do not reach a point where the core is expected to begin
propagating its accumulated heat as a luminosity wave. We also include the treatment of sink particle mergers used in \citetalias{Prole2022}.

For primordial chemistry, there is no first core as seen in present-day star formation. Fragmentation is expected to ceases once the collapse becomes adiabatic at $\sim 10^{-4}$ g cm$^{-3}$ (e.g. \citealt{Omukai2005}). Introducing sink particles at lower densities underestimates that degree of fragmentation in the disk (\citetalias{Prole2022}). Ideally, sink particles would be introduced at $10^{-4}$ g cm$^{-3}$ but this is currently computationally non-viable. For our purpose of following the gas fragmentation for $\sim 1000$ yr after sink creation, we find the highest viable resolution is $\sim 10^{-8}$ g cm$^{-3}$.

The accretion radius of a sink particle $R_{\text{sink}}$ is chosen to be the Jeans length $\lambda_{\text{J}}$ corresponding to the sink creation density. We set the minimum cell length to fit 16 cells across the sink particle in compliance with the Truelove condition, giving a minimum cell volume $V_{\text{min}}=(R_{\text{sink}}/8)^3$. The minimum gravitational softening length for cells and sink particles $L_{\text{soft}}$ is set to $R_{\text{sink}}/8$. Calculating $\lambda_{\text{J}}$ requires an estimate of the gas temperature, which we take from \citetalias{Prole2022}. The values of $R_{\text{sink}}$, $V_{\text{min}}$  and $L_{\text{soft}}$ are given in table \ref{table:sink}.

\begin{table}
	\centering
	\caption{Parameters for the simulations: sink particle creation density, sink radius radius, minimum cell volume and minimum gravitational softening lengths.}
	\label{table:sink}
	\begin{tabular}{lccr} 
		\hline
		$\rho_{\text{sink}}$ [g cm$^{-3}$] &  $R_{\text{sink}}$ [cm] & $V_{\text{min}}$ [cm$^{3}$] & $L_{\text{soft}}$ [cm]\\
		\hline
		$10^{-10}$ &  1.37$\times 10^{14}$ & 5.10$\times 10^{39}$ & 1.72$\times 10^{13}$\\
		$10^{-9}$ &  4.56$\times 10^{13}$ & 1.86$\times 10^{38}$ & 5.70$\times 10^{12}$\\
		$10^{-8}$ &  1.53$\times 10^{13}$ & 6.95$\times 10^{36}$ & 1.91$\times 10^{12}$\\

		\hline
	\end{tabular}
\end{table}

\section{Simulations}
\label{sec:sims}
\subsection{Non-magnetised collapse}
\label{sec:nonmag}
Initially, we assume that the magnetic field is too weak to affect the collapse significantly (see Section \ref{sec:comparison}) and run a pure hydrodynamic collapse of primordial gas. The initial conditions consist of a Bonner-Ebert sphere \citep{Ebert1955,Bonnor1956} categorised by central density $\rho_c$=2$\times$10$^{-20}$ g cm$^{-3}$ and radius R$_{\text{BE}}$=1.87 pc, placed in a box of side length 4R$_{\text{BE}}$ and temperature 200K. The stable Bonner-Ebert density profile was enhanced by a factor of 2 to promote collapse, such that the new central density was 4$\times$10$^{-20}$ g cm$^{-3}$. The initial abundances of H$_2$, H$^{+}$, D$^{+}$ and HD are x$_{\text{H}_{2}}$=10$^{-3}$, x$_{\text{H}^{+}}$=10$^{-7}$, $x_{\text{D}^{+}}$=2.6$\times$10$^{-12}$ and $x_{\text{HD}}$=3$\times$10$^{-7}$. A random velocity field was imposed on the box, generated from the Burgers turbulent power spectrum $\propto k^{-2}$ \citep{Burgers1948}. The rms velocity was scaled to give a ratio of kinetic to gravitational energy $\alpha$=0.05. The simulations were performed with a refinement criterion that ensured that the Jeans length was always resolved by at least 32 cells, as required by the findings of \cite{Federrath2011}. The simulations were repeated for 3 different velocity fields, which we henceforth refer to as seeds A, B and C. We repeat the simulations increasing the sink particle creation densities from 10$^{-10}$ to 10$^{-8}$ g cm$^{-3}$ to make sure the findings aren't resolution dependent.

\begin{table}
	\centering
	\caption{For each simulation parametrised by its seed field and sink creation density, we give the rms magnetic field strength and ratio of magnetic to gravitational energy within the MHD zoom-in simulations.}
	\label{table:mag}
	\begin{tabular}{cccc} 
		\hline
		Seed & $\rho_{\text{sink}}$ [g cm$^{-3}$]  &  B$_{\text{rms}}$ [mG] & E$_{\rm B}$/E$_{\rm grav}$\\
		\hline
		A & $10^{-10}$   &  6.94 & 0.38\\
		A & $10^{-9}$   &  6.90 & 0.37\\
		A & $10^{-8}$   & 6.94 & 0.37\\
		B & $10^{-10}$   &  11.74 & 0.25\\
		B & $10^{-9}$   &  11.68 & 0.25\\
		B & $10^{-8}$  &  11.78 & 0.25\\
		C & $10^{-10}$   &  7.42 & 0.40\\
		C & $10^{-9}$   &  7.40 & 0.41\\
		C & $10^{-8}$  &  7.48 & 0.41\\
		
		\hline
	\end{tabular}
\end{table}

\begin{figure*}
	 \hbox{\hspace{-0.5cm} \includegraphics[scale=0.8]{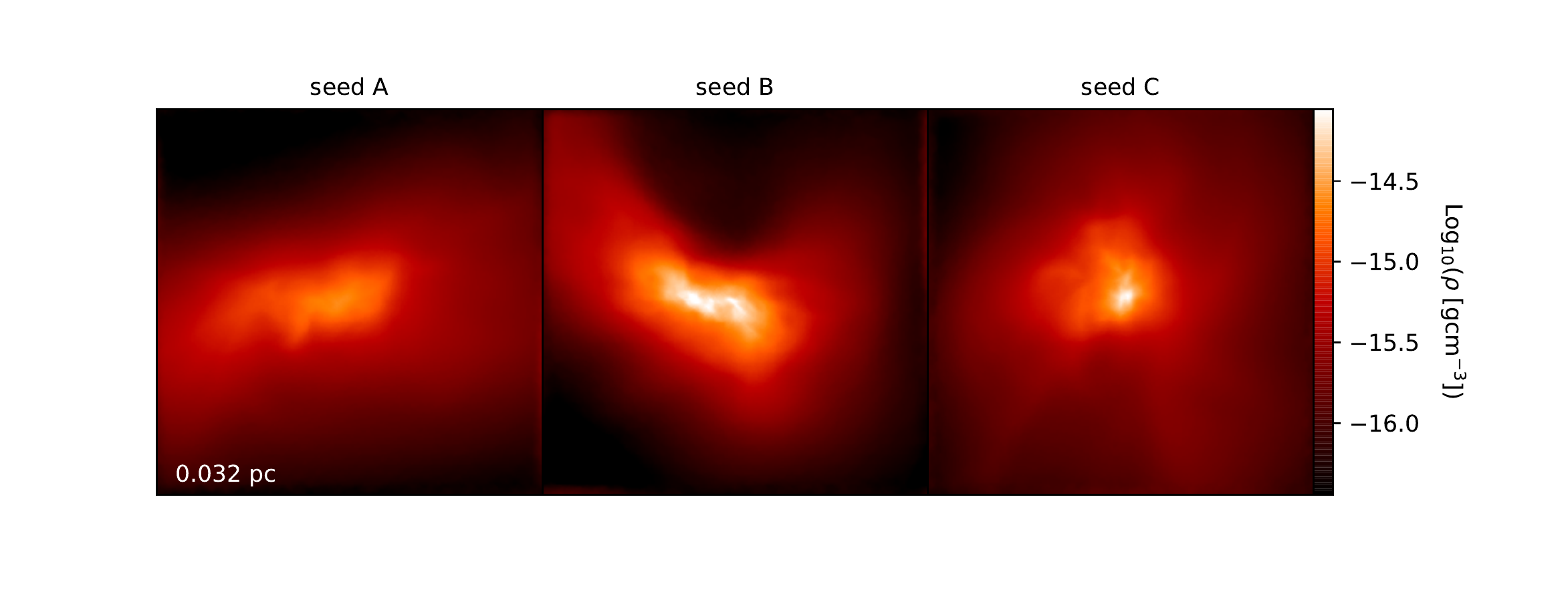}}
    \caption{Average density within the inner 0.032 pc of the non-magnetised collapse, once the central density has reached $\sim$10$^{-13}$ g cm$^{-3}$, projected onto 500$^3$ cubes and flattened along the y axis. These serve as the initial conditions for the magnetised stage of the collapse.}
    \label{fig:ics}
\end{figure*}

\begin{figure}
	 \hbox{\hspace{-0.5cm} \includegraphics[scale=0.6]{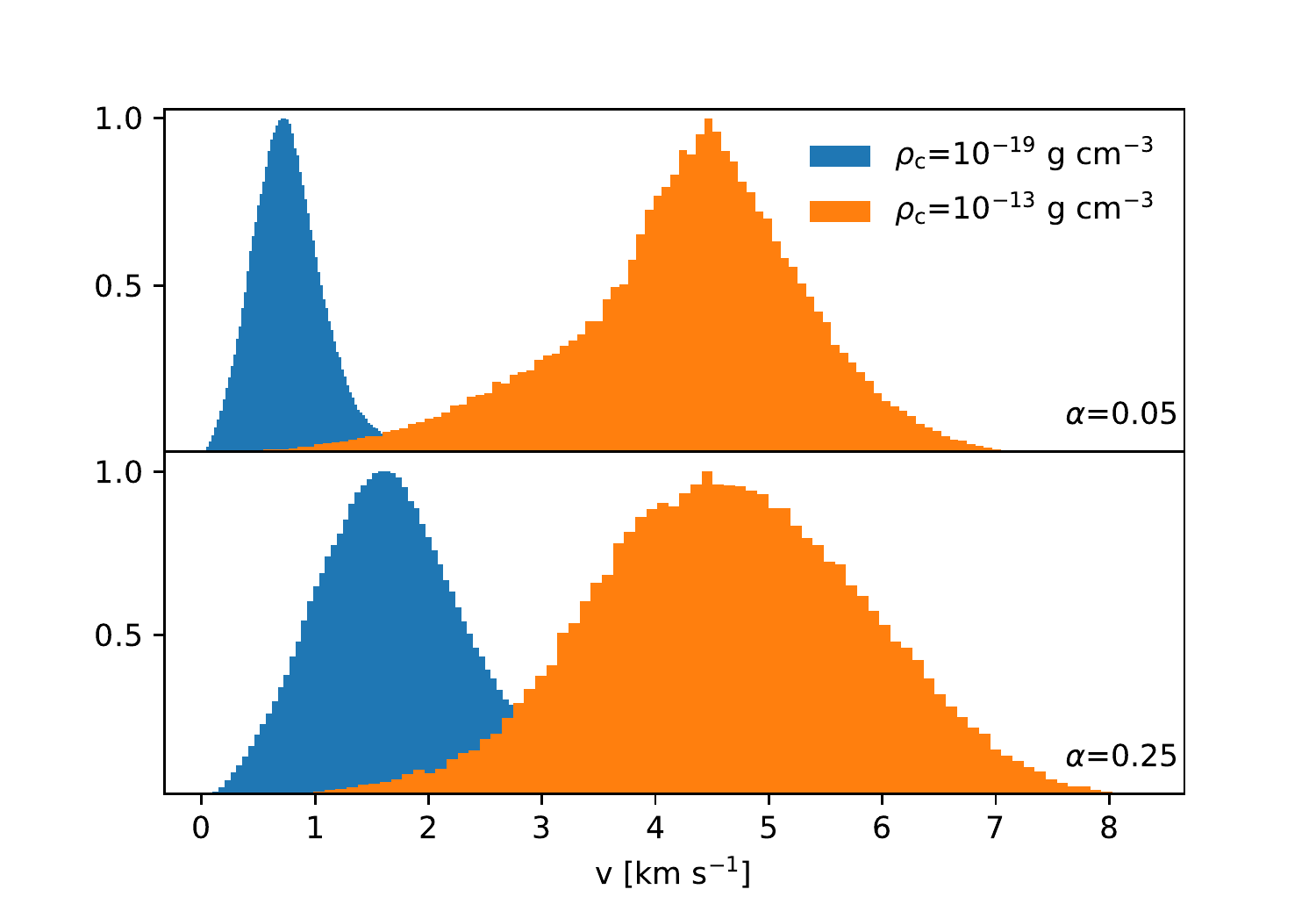}}
    \caption{Normalised, mass weighted histograms of the initial velocity field and the field within the central 0.032pc when the central density reaches $\sim$10$^{-13}$ g cm$^{-3}$, for the $\alpha$=0.025 and $\alpha$=0.005 runs of \citetalias{Prole2022}.}
    \label{fig:oldvels}
\end{figure}

\subsection{Magnetised collapse}
\label{sec:zooms}
We allow an initial non-magnetised collapse to run until the central density reaches $\sim$10$^{-13}$ g cm$^{-3}$, before cropping the central region and applying a saturated magnetic field. At this point, the smallest cells are of scale $\Delta x \sim 14.5$AU. The cropped box size is chosen by calculating the radius which corresponds to a free-fall time of 10$^4$ yr, ensuring that further collapse of the order of 1000 yr can be safely followed within the new box size. The new box size was calculated to be 0.032pc and has periodic boundary conditions. The cropped boxes are shown in figure \ref{fig:ics}. To generate the saturated magnetic fields, we used the Kazantsev $P_k \propto k^{3/2}$  power spectrum with 200 modes (and 200 negative modes), resulting in a cube with spatial resolution $\Delta x \sim 17$AU, such that the field is resolved in the central region of the collapse. The imposed magnetic field is scaled to the maximum strength by assuming that the magnetic energy has saturated with the velocity field. While it might be expected that saturated magnetic field strength depends on the initial velocity field strength, Figure \ref{fig:oldvels} shows the velocity fields from the $\alpha$=0.025 and $\alpha$=0.005 runs in \citetalias{Prole2022} within the central 0.032pc, when the central density reaches 10$^{-13}$ g cm$^{-3}$. The strength of the central velocity field does not depend on the initial field strength, hence the saturated magnetic field strength does not depend on the initial velocity field. The method of generating a 3D field from the 1D power spectrum ($P_k \propto k^{3/2}$) is as follows; the 1D spectrum was converted to 3D $k$-space, where $k$ is the number of cycles per boxsize, spanning from 0-N where N is the number of modes. The energies of each $k_{\alpha,\beta,\gamma}$ coordinate are given by dividing $P_k$ by the shell volume $4\pi|k_{\alpha,\beta,\gamma}|^2dk$ to give $P_{\alpha,\beta,\gamma}$. These energies are the squared amplitudes of the waves i.e. $\sqrt{P_{\alpha,\beta,\gamma}}=|B_{\alpha,\beta,\gamma}|$, which are split between the 3 spatial dimensions to give a vector $\vec{B}_{\alpha,\beta,\gamma}$. Spatial dimensions split the energy equally, each having amplitude $\sqrt{1/3 \ P_{\alpha,\beta,\gamma}}$. Gauss’s law for magnetism states that the divergence of magnetic fields is 0, so we remove the compressive components of the modes by subtracting the longitudinal component of the amplitude $\hat{k}\frac{(\vec{k} \cdot \vec{B})}{|\vec{k}|}$. The modes are given random phase offsets $\phi_{\alpha,\beta,\gamma}$ as 

\begin{equation}
B_{\alpha,\beta,\gamma}=|B_{\alpha,\beta,\gamma}|[\cos(\phi_{\alpha,\beta,\gamma})+\sin(\phi_{\alpha,\beta,\gamma})i]
\end{equation}
and converted into 3D real space via inverse Fourier transform. The field is then interpolated onto the {\sc Arepo} cell coordinates and rescaled to give the desired rms strength. The rms strength is chosen so that the magnetic field energy reaches equipartition with the velocity field. Magnetic energy density is given by $B^2/8\pi$, so we scale the rms magnetic field strength as $B_{\text{rms}}=\sqrt{8\pi \epsilon_{\text{KE}}}$, where $\epsilon_{\text{KE}}$ is the volumetric kinetic energy density.\\

We repeat the simulations with no magnetic fields as a control case. For seed A, we repeat the simulations with a uniform field with the same rms field strength as the $k^{3/2}$ field, to show the effects of under resolving the field in the initial conditions. We also repeat the $\rho_{\text{sink}}=10^{-13}$ g cm$^{-3}$ version of seed A with a less restrictive refinement criterion that requires only 16 cells per Jeans length, to compare the dynamo amplification to the 32 cells per Jeans length case. This gives a total of 22 zoom-in simulations from the initial 3 full-scale simulations. The initial conditions for the magnetised collapses are summarised in table \ref{table:mag}.

\begin{figure}
\begin{minipage}[b]{0.55\linewidth}
  \centering
  \hbox{\hspace{-0.5cm}\includegraphics[scale=0.55]{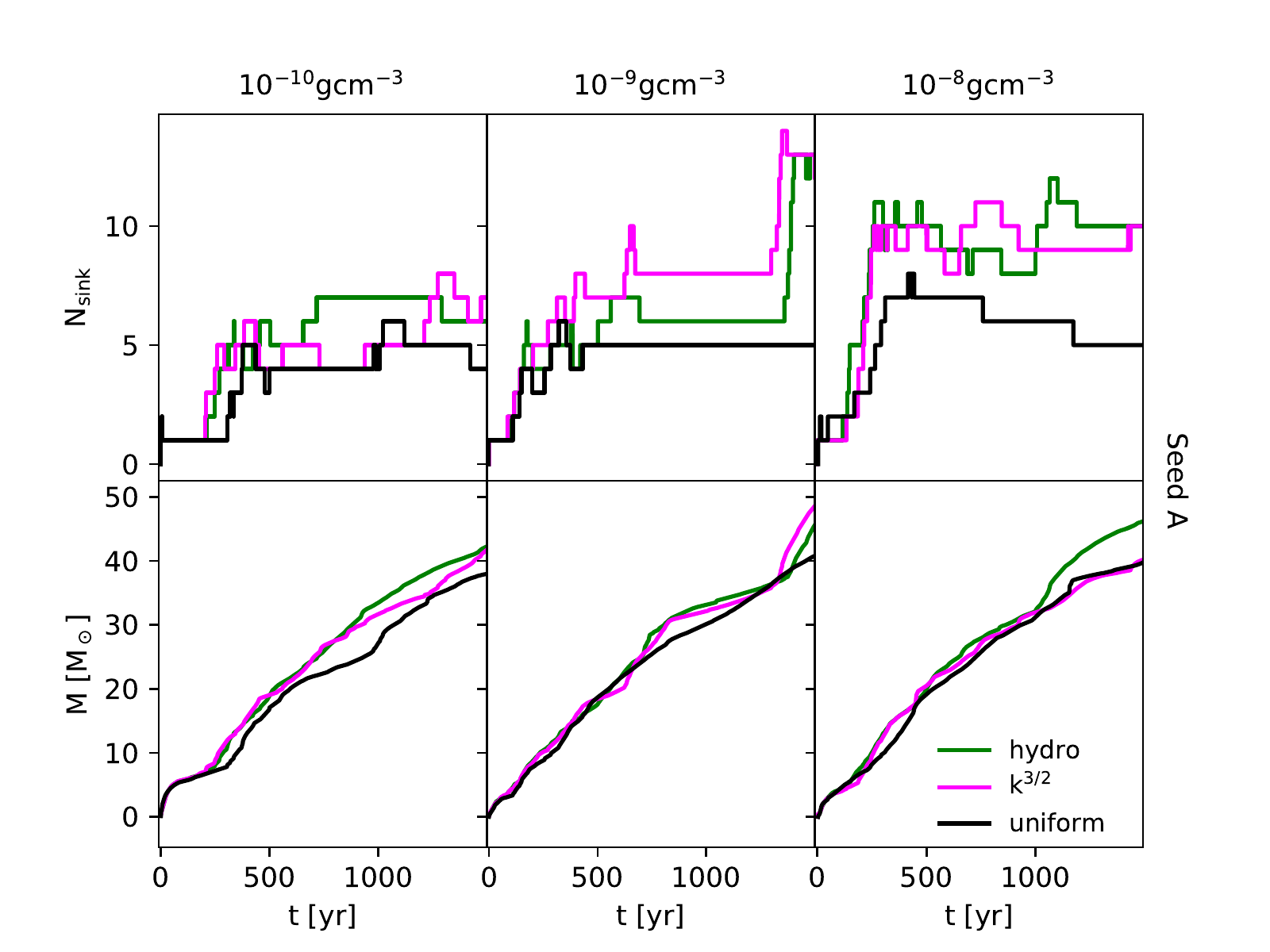}}
  
\end{minipage}
\hfill
\begin{minipage}[b]{0.45\linewidth}
  \centering
  \hbox{\hspace{-0.5cm}\includegraphics[scale=0.55]{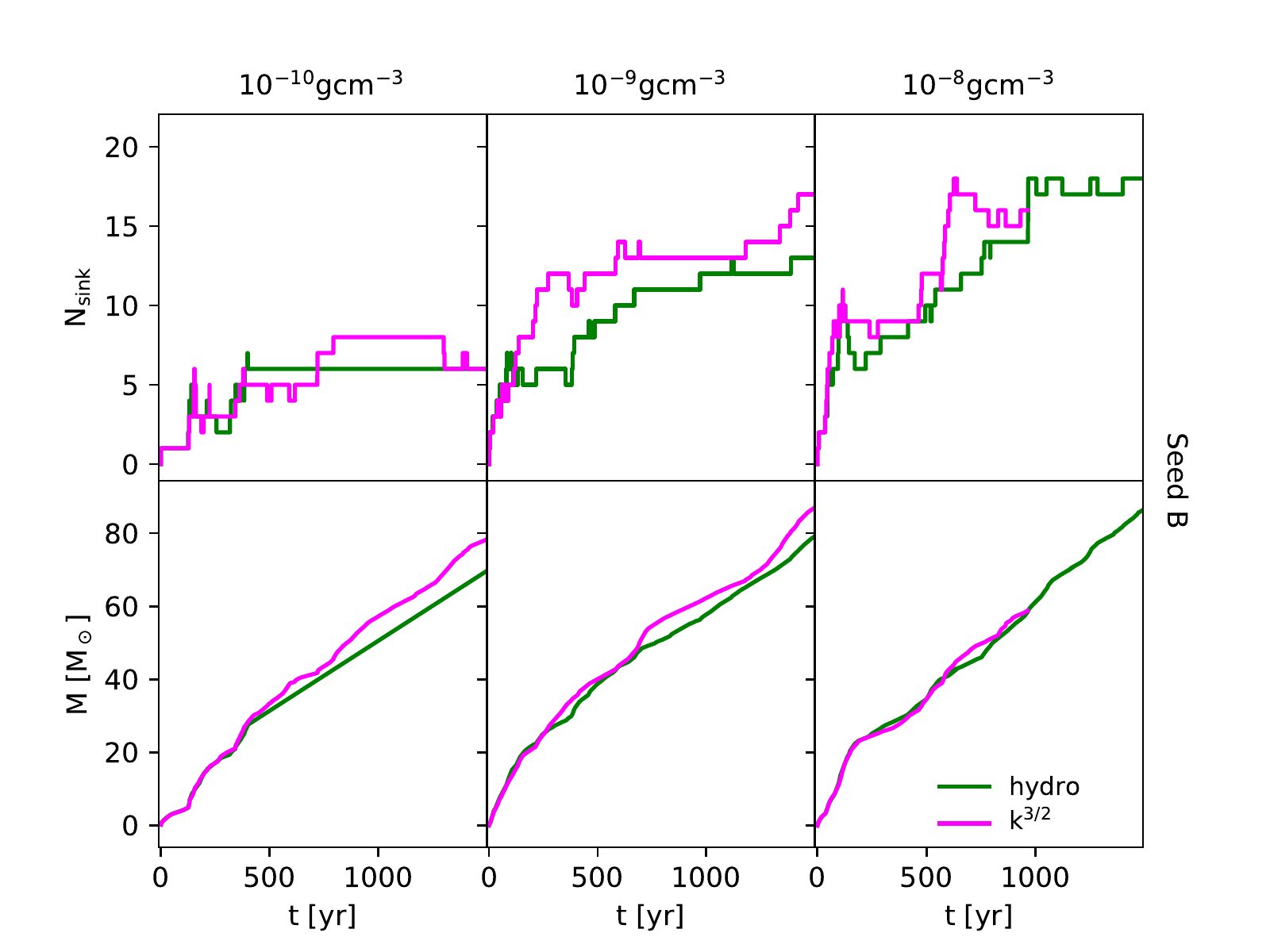}}
 
\end{minipage}
\hfill
\begin{minipage}[b]{0.45\linewidth}
  \centering
  \hbox{\hspace{-0.5cm}\includegraphics[scale=0.55]{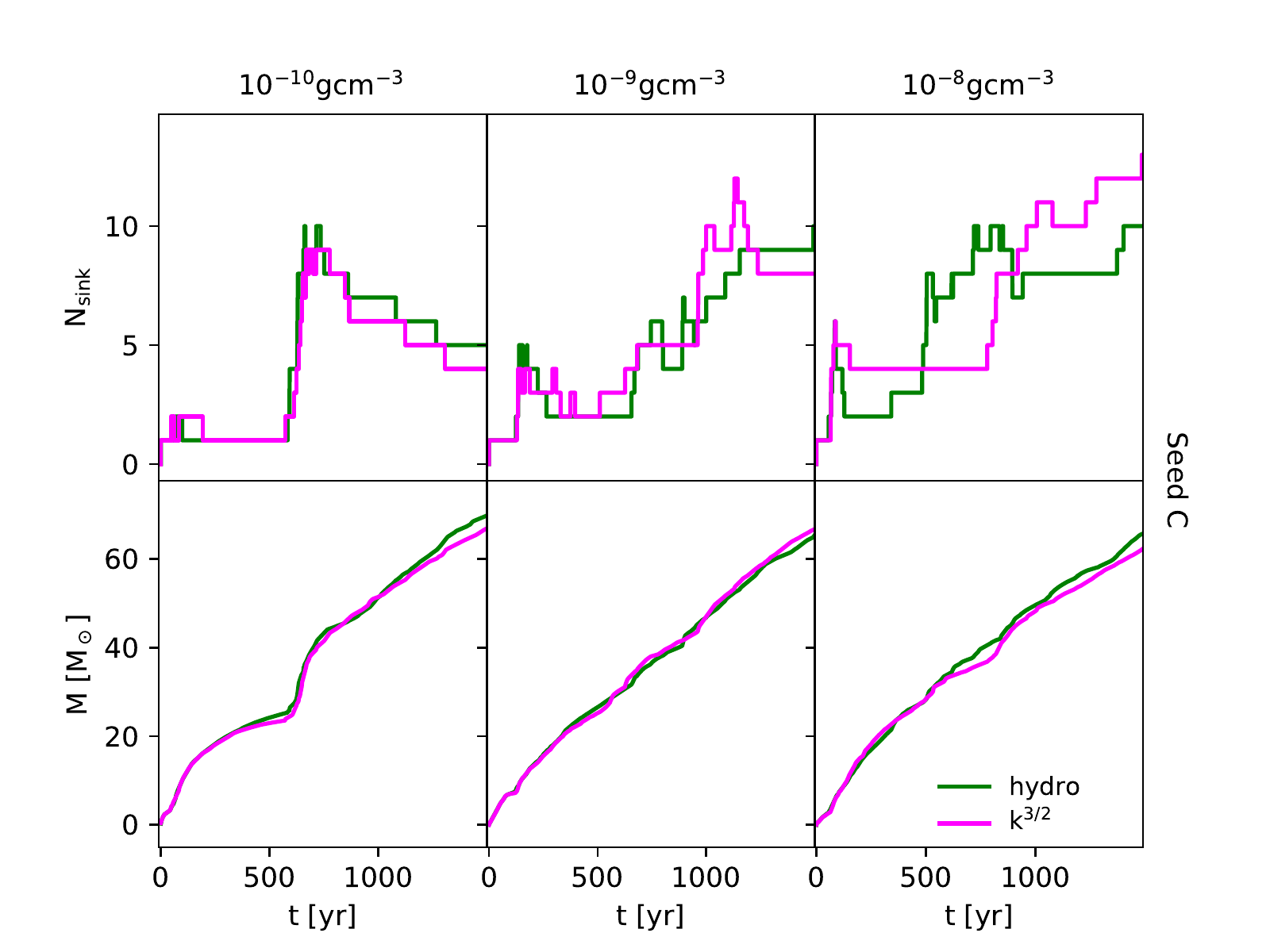}}
  
\end{minipage}

 \caption{Number of sink particles formed and total mass in sink particles as a function of time, for the 3 initial velocity and magnetic fields. t = 0 corresponds to the formation time of the first sink particle formed. The sink particle creation density is indicated above the columns.}
 \label{fig:sinks}
\end{figure}

\begin{figure*}
	 \hbox{\hspace{0cm} \includegraphics[scale=0.7]{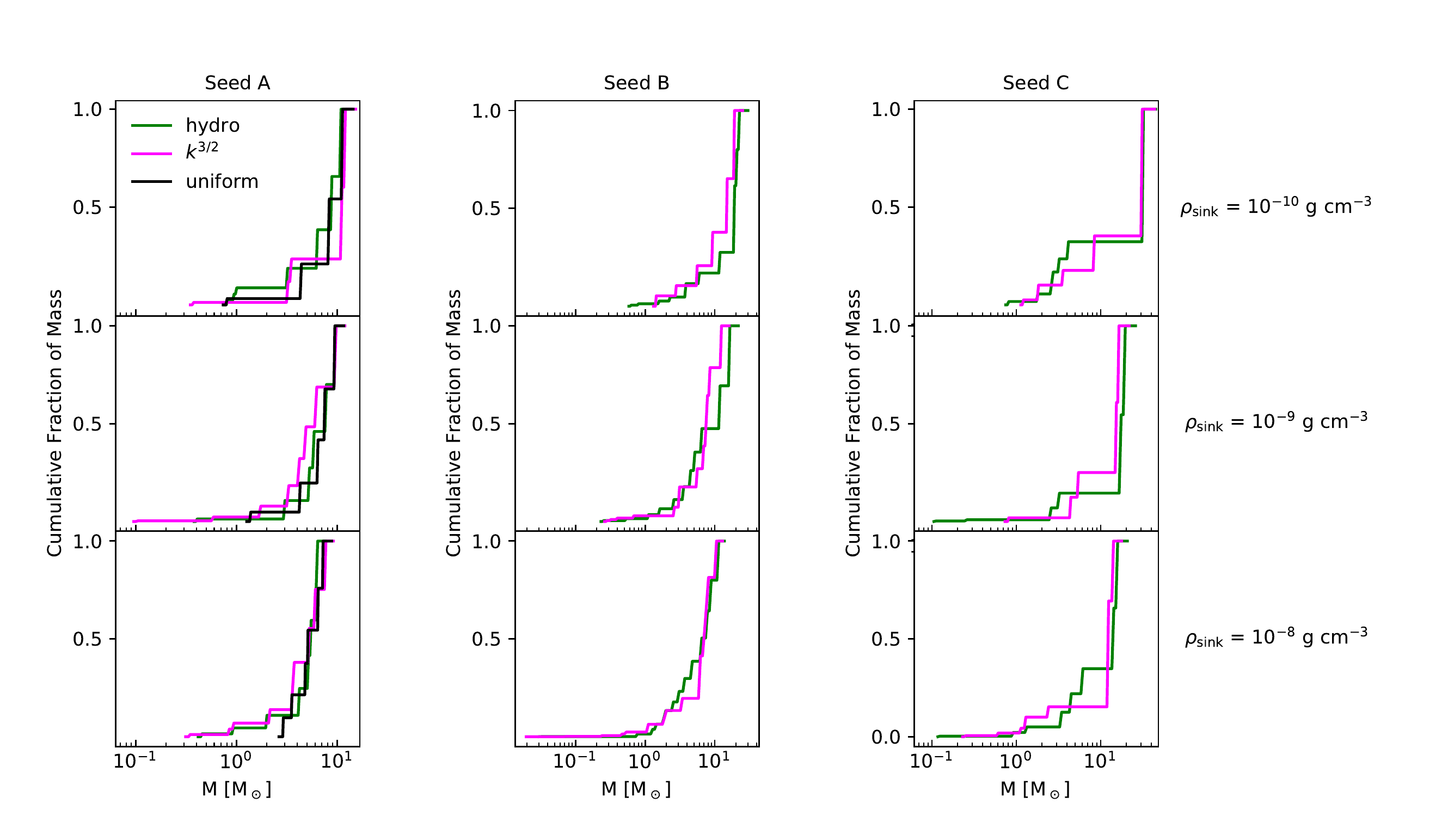}}
    \caption{Normalised cumulative IMFs for the hydro and $k^{3/2}$ runs at $\sim$ 1000 yr after the formation of the first sink particles. The sink particle creation densities used are shown on the right side of the figure.}
    \label{fig:IMF}
\end{figure*}

\begin{figure*}
	 \hbox{\hspace{-3cm} \includegraphics[scale=0.68]{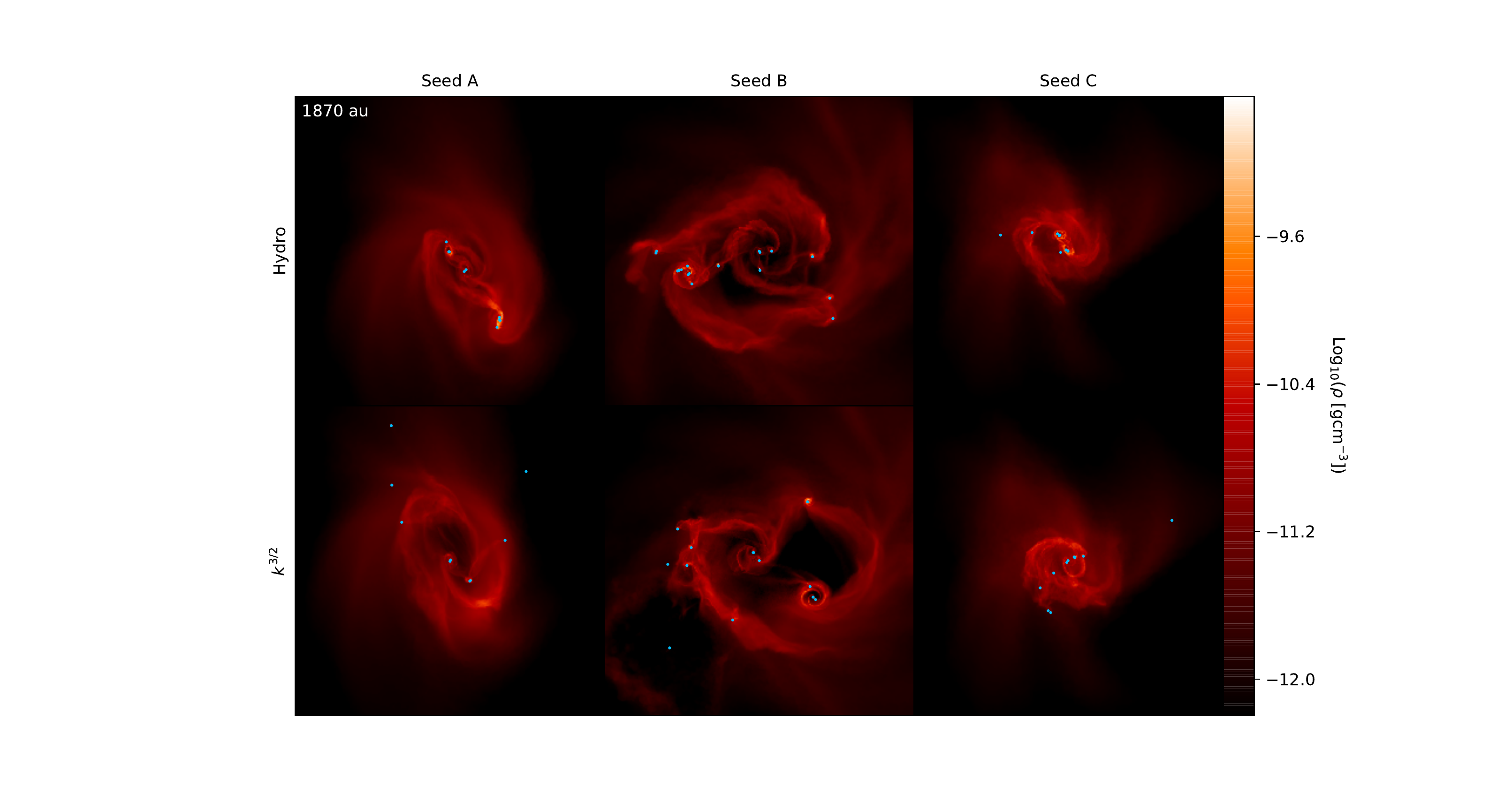}}
    \caption{Density projections for the highest resolution ($\rho_{\text{sink}} = 10^{-8}$ g cm$^{-3}$) runs, comparing the non-MHD (top) and $k^{3/2}$ magnetic field (bottom) scenarios. Sink particles are shown as blue dots.}
    \label{fig:projections}
\end{figure*}

\begin{figure*}
	 \hbox{\hspace{-3cm} \includegraphics[scale=0.68]{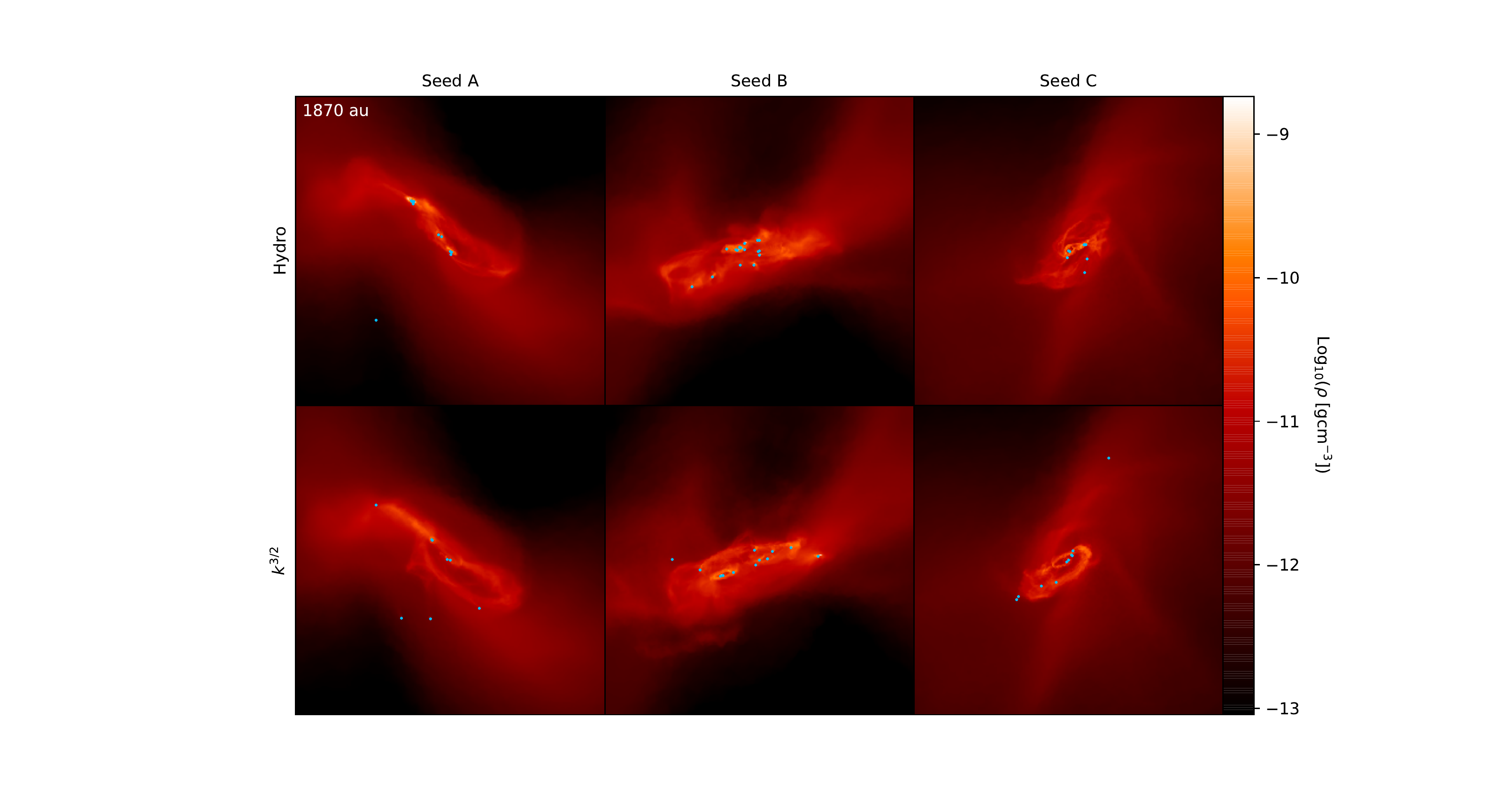}}
    \caption{The same as Figure \ref{fig:projections}, but showing the edge on perspective of the discs.}
    \label{fig:projections2}
\end{figure*}

\section{Fragmentation behaviour}
\label{sec:frag}
For the 3 seed fields, Figure \ref{fig:sinks} shows the number of sink particles and total mass in sink particles as a function of time, for increasing sink particle creation density. For all 3 seeds, the number of sink particles is not significantly lower in the magnetised case compared to the hydrodynamic cases. There is therefore no evidence of reduced fragmentation owing to magnetic support. While the pause in sink formation seen in the top left panel of seed C could be interpreted as delayed star formation due to magnetic fields, the total mass accreted onto sink particles is unaffected by the fields. This indicates that the differences are due to the stochastic nature of N-body simulations. In the hydrodynamic case, increased fragmentation with increasing sink creation density occurs as described in \citetalias{Prole2022}, despite the employed zoom-in method described in Section \ref{sec:zooms}. The presence of magnetic fields does not change this outcome within the range of resolutions tested. The cumulative IMFs at $\sim 1000$ yr after sink formation are shown in Figure \ref{fig:IMF} and reveal no trends between the hydrodynamic and MHD runs. The systems at $\sim 1000$ yr are shown in Figure \ref{fig:projections}. While the scale and orientation of the systems are unaffected by the magnetic fields, there are some structural differences which are to be expected from stochastic N-body simulations.

When the $k^{3/2}$ field was replaced with a uniform field for velocity seed A, the system does experience decreased fragmentation as the resolution is increased. The fragmentation behaviour possibly converges with resolution in this case, although that is impossible to interpret with only 1 realisation of the fields. Decreased fragmentation in the presence of initially uniform magnetic fields has been seen in many previous  present-day star formation studies (e.g. \citealt{Price2007,Machida2013}). This effect is likely due to a mixture of magnetic tension transferring angular momentum away via magnetic breaking and pressure support against Jeans instabilities. The lesser effect of the field at lower resolutions is due to the gas's tendency to fragment less at lower resolutions in both the hydrodynamic and MHD scenarios.

Suppressed fragmentation by uniform fields at these scales reveals a non-obvious danger of including a tangled magnetic field in the initial conditions as has been done in previous studies (e.g. \citealt{Sharda2020}). For a given simulation box, only a small region collapses to form stars, so an insufficiently resolved $k^{3/2}$ magnetic field can act as a uniform field across the relevant areas in the early collapse, providing false support against fragmentation. This may not be a problem in the later collapse as the small scale turbulent dynamo should convert the field into a $k^{3/2}$ power spectrum, but the early effects of the field could significantly alter the gas flow into the star-forming region. Our high resolution $k^{3/2}$ fields were introduced across a small region of the box once relevant areas of star formation were revealed, so we are confident that the gas sees a $k^{3/2}$ field (see Section \ref{sec:comparison}).

\section{Field behaviour}
\label{sec:field}
\begin{figure}
	 \hbox{\hspace{-0.5cm} \includegraphics[scale=0.5]{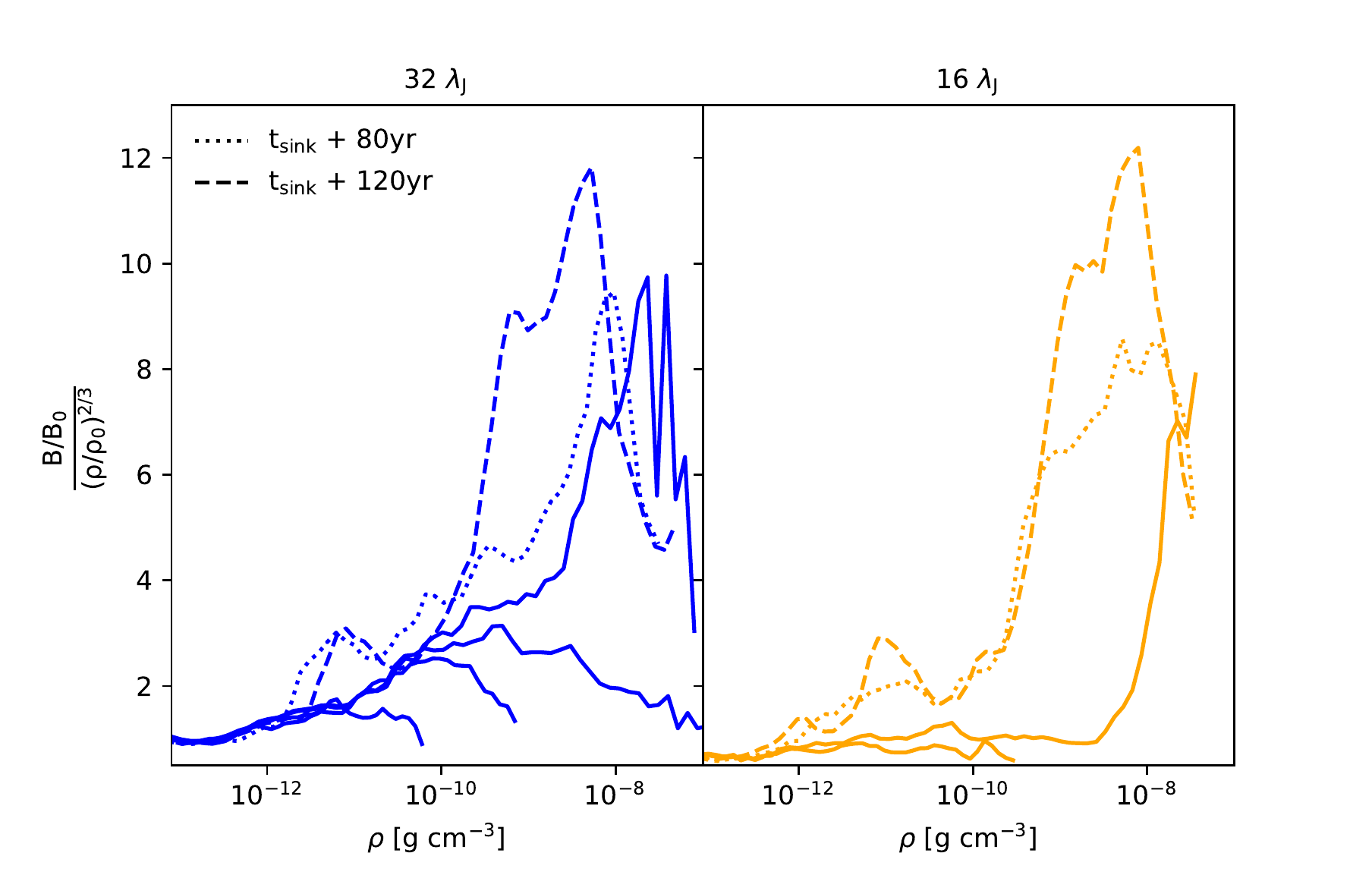}}
    \caption{Comparison of the dynamo amplification between the 32 and 16 cells per Jeans length runs for seed field A, for the highest sink particle creation density. The field strength is normalised by $\rho^{2/3}$ i.e. the growth due to flux-freezing. Any residual positive gradient is due to dynamo amplification. The solid lines show the field at different central densities leading up to and including sink particle formation, while the dotted and dashed lines show the field at 80 and 120 yr after sink formation, respectively. }
    \label{fig:amplification}
\end{figure}

Figure \ref{fig:amplification} shows the field strength normalised by $\rho^{2/3}$, which is the amplification expected due to flux-freezing alone. Any positive gradients in the resulting curve correspond to dynamo amplification. The runs with 32 cells per Jeans length  do experience dynamo action while the central density increases towards the maximum density of the simulations, unlike the run with 16 cells per Jeans length. This aligns with the findings of \cite{Federrath2011}. The amplification is a factor $\sim 3$ in the densest regions. However, after sink particle formation the amplification rises rapidly to $\sim 10$ in both cases.

\begin{figure*}
	 \hbox{\hspace{1cm} \includegraphics[scale=0.6]{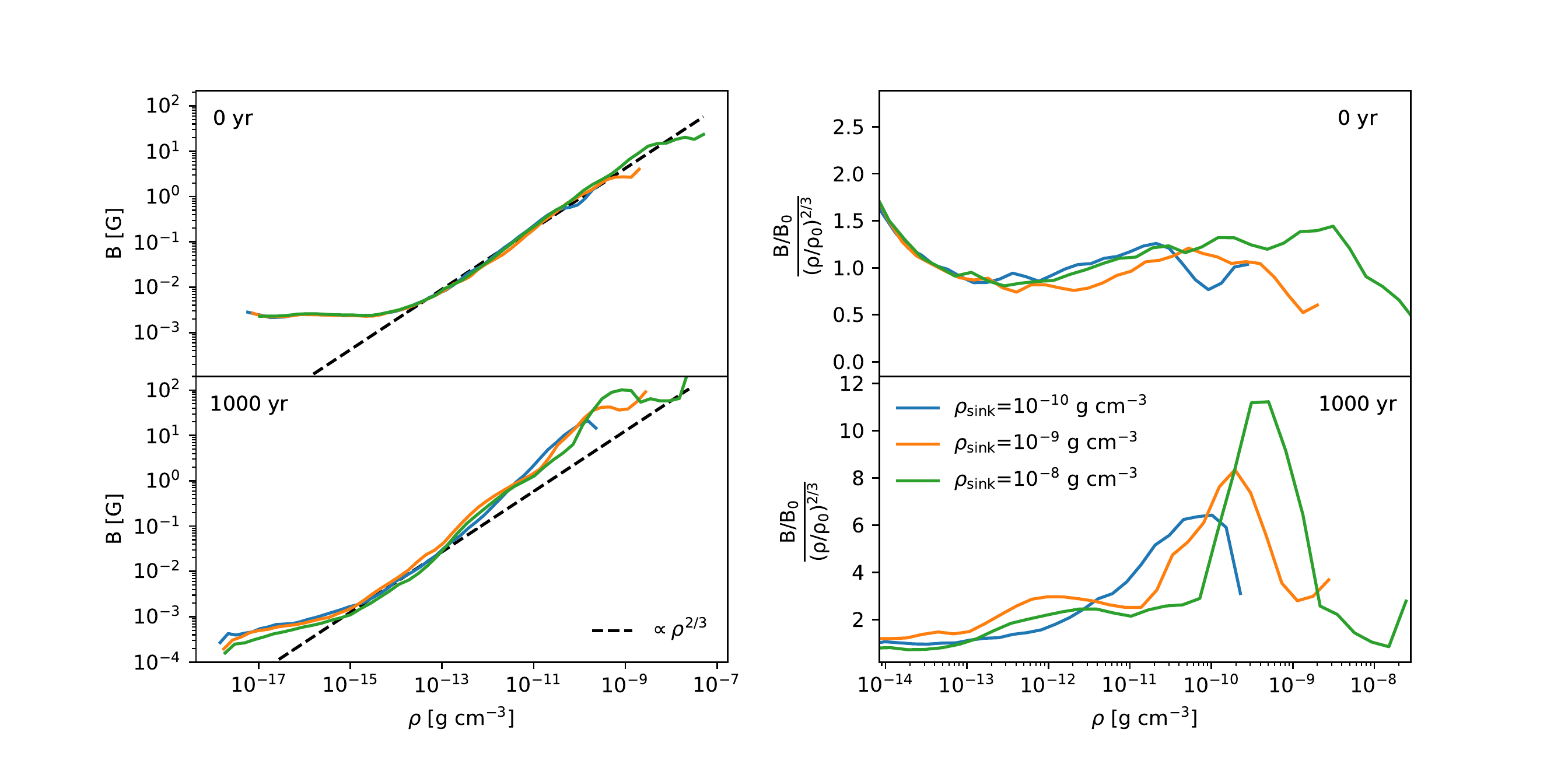}}
    \caption{Left: volume-weighted magnetic field strength as a function of density for the three maximum resolutions considered from seed field C, shown at a time just before the formation of the first sink particle (top) and 1000 yr later (bottom). The dashed line shows the field growth expected from flux freezing. Right: Field strengths divided by the flux-freezing line, such that growth above values of 1 indicate dynamo amplification.}
    \label{fig:jeans}
\end{figure*}

The field strength just before the formation of sink particles is shown for the different resolutions used in Figure \ref{fig:jeans}. Higher resolutions allow for higher maximum densities and hence stronger maximum magnetic field strengths due to flux-freezing. By normalising by $\rho^{2/3}$, it is revealed that increasing the maximum refinement level (and decreasing the gravitation softening length) only affects the dynamo amplification at the highest available densities (smallest scales), where the amplification gets progressively stronger.

\begin{figure}
	 \hbox{\hspace{-0.5cm} \includegraphics[scale=0.6]{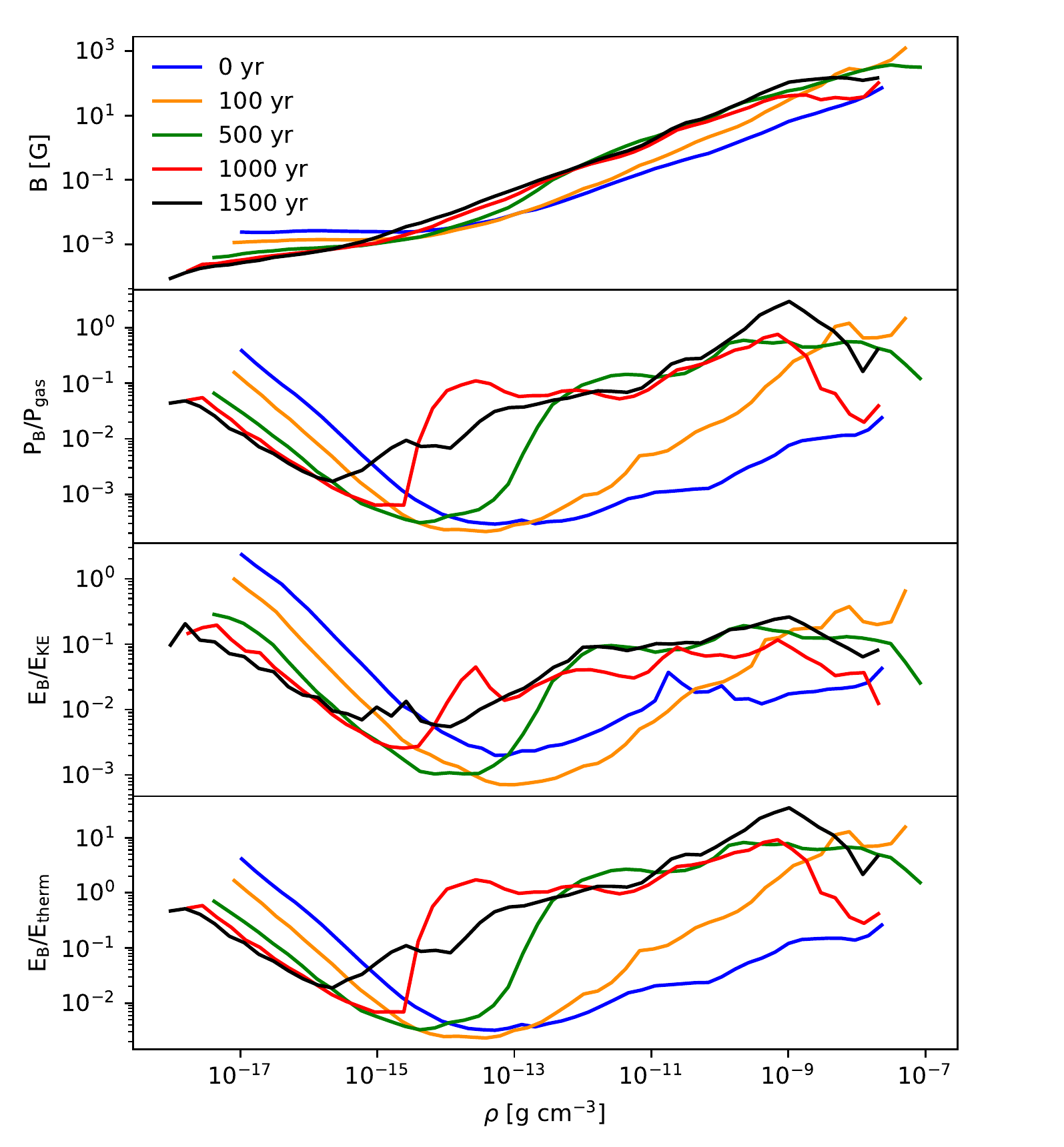}}
    \caption{Volume-weighted density profiles for seed field C of the magnetic field strength, ratio of magnetic to thermal pressure, ratio of magnetic to kinetic energy and ratio of magnetic to thermal energy.}
    \label{fig:field}
\end{figure}

Figure \ref{fig:field} shows the evolution of field quantities as a function of density, for the $10^{-8}$ g cm$^{-3}$ run. The field strength within the entire collapsing region grows until seemingly converging after 1000 yr. For our small-scale fields, we do not expect magnetic tension will play an important role. The magnetic pressure however is isotropic and in theory can support against fragmentation. The magnetic and thermal pressure are given by 

\begin{equation}
P_{\text{B}} = \frac{B^2}{8 \pi},
\label{eq:pressures}
\end{equation}

\begin{equation}
P_{\text{gas}} = \frac{k_{\text{B}} \rho T}{\mu m_p}
\label{eq:pressures2}
\end{equation}

respectively, where $k_{\text{B}}$ is Boltzmann's constant, T is the temperature, $\mu$ is the mean molecular weight and m$_{\text{p}}$ is the mass of a proton. The magnetic pressure is significantly lower than the thermal pressure at low densities and high densities pre-sink formation, explaining why it was unable to suppress fragmentation. Amplification of the field strength leads to the magnetic and thermal pressures becoming comparable at densities of $10^{-10}$ g cm$^{-3}$ and upwards by the end of the simulation. Additionally, the maximum density of the simulations was reached before magnetic pressure could grow to become the dominant term. The magnetic pressure grows at minimum as $\rho^{4/3}$ due to flux-freezing, while the gas pressure has a linear dependency on density, so the gas pressure likely becomes dominated by the magnetic contribution at higher densities than this investigation has resolved. As fragmentation is expected to occur on smaller Jeans scales with increasing density, this dominant magnetic pressure may suppress fragmentation on the smallest scales between $10^{-10}-10^{-4}$ g cm$^{-3}$ before the adiabatic core forms. However, one-zone calculations with an added protostellar model for zero metalicity stars by \cite{Machida2015} suggest a rapid increase in temperature at $10^{-6}$ g cm$^{-3}$, likely rendering the gas stable to fragmentation. This only leaves 2 orders of magnitude in density higher than our sink particle creation density where magnetic fields could suppress fragmentation.

The ratio of magnetic to kinetic energy settles at a value of $\sim 0.3$ in the highest density regions while the ratio of magnetic to thermal energies reaches $\sim 10$. This saturation value fits in line with previous turbulent box studies of the small-scale turbulent dynamo (e.g. \citealt{Haugen2004}, \citealt{Federrath2011a}, \citealt{Schober2015}).

\begin{figure}
	 \hbox{\hspace{-0.5cm} \includegraphics[scale=0.6]{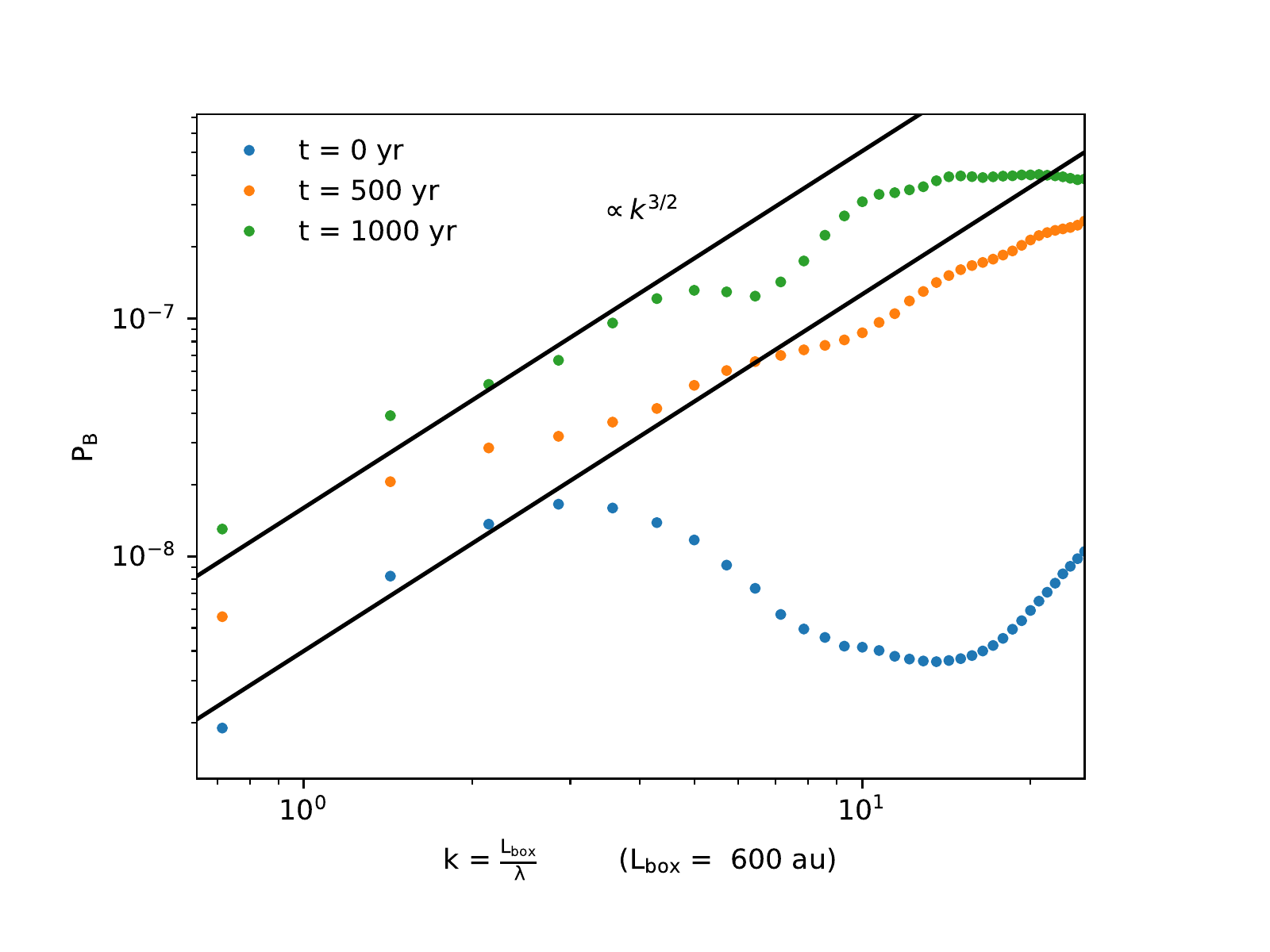}}
    \caption{The magnetic power spectra within the central 600 au at times 0, 500 and 1000 yr after the formation of the first sink, for seed field C. The highest modes shown on the plot correspond to scales of 25 au. The black lines show a $k^{3/2}$ power law.}
    \label{fig:power_spectrum_end}
\end{figure}

The evolution of the magnetic power spectrum within the central 600 au around the most massive sink is shown in Figure \ref{fig:power_spectrum_end}. These were created by projecting the {\sc Arepo} cells onto a 50$^3$ uniform grid, resolving scales of 12 au. The peak of the spectrum moves to smaller spacial scales after the formation of sink particles, following a $k^{3/2}$ power law down to scales of $\sim 120$ au at 1000 yr after the formation of the first sink particle. The energy continues to grow on smaller scales down to the Nyquist frequency of the projection, corresponding to $\sim 25$ au.

Our results suggest that the inclusion of magnetic fields is unimportant in numerical simulations of Pop III star formation up to densities of 10$^{-8}$ g cm$^{-3}$. For most previous Pop III studies, the resolution is too poor to capture the small-scale fragmentation that could be suppressed by magnetic fields at densities higher than explored in the simulations presented here (e.g. \citealt{Greif2011,Smith2011,Clark2011,Susa2014,Stacy2016,Wollenberg2019,Sharda2020}). This result renders the task of finding the primordial IMF through numerical simulations less computationally expensive and time consuming. 

\section{Comparison with previous studies}
\label{sec:comparison}
\cite{Machida2008a} produced jets in set-ups where the initial magnetic field was dominant over rotation ($\gamma_0 > \beta_0$) above a threshold magnetic field strength of $10^{-9}$ G at $10^{-21}$ g cm$^{-3}$. They did not report the evolution of magnetic field strengths at different central densities, so we can not compare the field strength at our starting density of $10^{-13}$ g cm$^{-3}$. Initial solid-body rotation was also not included in our initial conditions, but from our turbulent velocity field we calculate that the ratio of total magnetic energy to rotational kinetic energy is $\sim 2$ when we introduce the field to the cropped box, making our model magnetically dominated. Despite this, there is no evidence for jet launching in our simulations. This is due to our use of a realistic $P(k) \propto k^{3/2}$ magnetic field power spectrum, which results in misaligned field vectors and hence the misaligned directional force of magnetic tensions. Also, our inclusion of a turbulent velocity field promotes fragmentation of the disc, differentiating the set-up from the idealised rotating cloud used in \cite{Machida2008a} and \cite{Machida2013}.

\begin{figure}
	 \hbox{\hspace{-0.5cm} \includegraphics[scale=0.7]{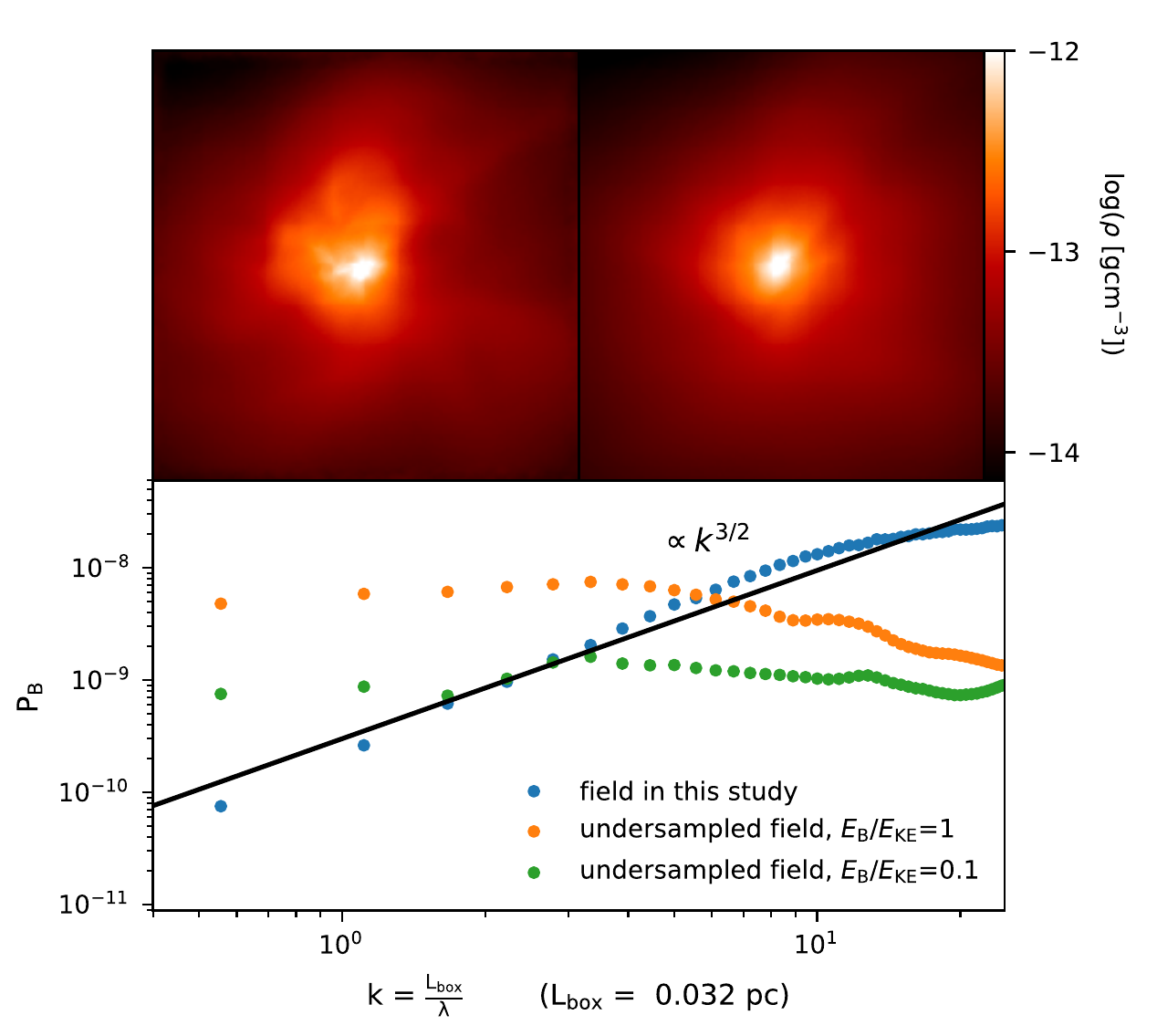}}
    \caption{Comparison of introducing a magnetic field late in the collapse versus introducing a low resolution field at the beginning of the collapse. Top - density projections of the non-magnetised (left) and $E_{\rm B}$/$E_{\rm KE}$=1 magnetised (right) clouds at the point when our tangled field is introduced as described in Section \ref{sec:zooms}. Bottom - Magnetic power spectrum of the field we interpolate onto the cells (blue) compared to the evolved spectra of the fields introduced at the beginning of the simulation at $E_{\rm B}$/$E_{\rm KE}$=1 (orange) and 0.1 (green). The black line shows a $k^{3/2}$ power law.}
    \label{fig:spectra}
\end{figure}

\cite{Sharda2020} used a $P(k) \propto k^{3/2}$ power spectrum, finding suppressed fragmentation and a reduction in the number of first stars with masses low enough that they might be expected to survive to the present-day. The magnetic field strengths at their maximum density of $10^{-11}$ g cm$^{-3}$ are similar to ours at the same density (1-10 G). However, they introduce their fields at a much lower density and hence poorer mesh resolution, and only populate the magnetic field with 20 modes throughout the 2.4 pc box. This raises concerns that the $k^{3/2}$ spectrum could be under-sampled, possibly allowing the magnetic field to be artificially uniform within the collapsing region. To demonstrate this, we repeat the initial stage of collapse described in Section \ref{sec:nonmag} with an initial $P(k) \propto k^{3/2}$ magnetic field consisting of 20 modes, scaled to $E_{\rm B}$/$E_{\rm KE}=0.1$ to replicate the strong field case in \cite{Sharda2020}, giving an rms field strength of 1.01 $\rm \mu$G. We repeat the simulation with a ratio of 1 to compare with the method from this paper, giving a field strength of 3.18 $\rm \mu$G. We allow the gas to collapse to the point where our non-magnetised box was cropped and our magnetic field was interpolated onto the {\sc Arepo} cells. Figure \ref{fig:spectra} compares the gas structure and power spectra of the fields at that point within the central 0.032pc cropped box. Our freshly introduced field is roughly $P(k) \propto k^{3/2}$ while the evolved fields have almost flat power spectra, indicating equal energies on all scales, which is not predicted by Kazantsev theory. The top panels of Figure \ref{fig:spectra} compare the gas structure of our magnetised initial conditions with the $E_{\rm B}$/$E_{\rm KE}$=1 simulation described above. The lack of significant difference in the structure supports our assumption that the field does not affect the collapse of the gas significantly before the point where we introduce our magnetic field. We attribute the lack of magnetic support in our study compared to \cite{Sharda2020} to our improved method of sampling the tangled $P(k) \propto k^{3/2}$ spectrum,  along with our increased maximum densities and spatial resolution allowing fragmentation on smaller Jeans scales.

\section{Caveats}
\label{sec:caveats}
The weak magnetic seed fields undergo amplification via the small-scale magnetic dynamo, during the early stages of collapse. In our initial non-magnetised simulations, we have assumed that the magnetic fields are too weak to affect the collapse, and only account for magnetic effects once the field is expected to have saturated. The magnetic fields generated from the $k^{3/2}$ spectrum are random, whilst the fields resulting from a fully magnetised treatment would be structured by the collapsing density distribution and velocity field. Furthermore, while the initial conditions were set up to a field energy at equipartition with the velocity field (i.e. their theoretical maximum strength and hence maximum support against fragmentation), Figure \ref{fig:field} clearly shows that the equipartition did not survive the early stages of collapse. We suspect that this is due to averaging the energies over the whole simulation box and spreading the field energy equally over entire volume, most of which did experience collapse. However, previous studies have found  saturation ratios of the dynamo in the range of $\sim 0.01 - 0.6$, which is in line with the fields in this study.

Our method of calculating the rms field strength was dependant on the mean kinetic energy in the cropped simulation box, which is sensitive to the size of the cropped box. Zooming into a smaller region around the center of collapse gives a higher average kinetic energy and hence a higher rms magnetic field strength. Our cropped box size was chosen to provide a large enough cloud to allow collapse for sufficient time without causing issues with the periodic boundary conditions.

Although we remove compressive modes from the magnetic fields generated from the $k^{3/2}$ power spectrum, $\bigtriangledown \cdot \vec{B}$ errors are produced when the field is interpolated onto the {\sc Arepo} non-uniform mesh. During the early stages of collapse, these errors are cleaned as described in Section \ref{sec:MHDarepo}. Our simulations also suffer from numerical diffusion. The combination of these effects results in a loss of magnetic energy in the early stage of collapse. While these issues would be reduced by implementing a magnetic field with fewer modes i.e. neglecting the smallest spatial scales, this is the equivalent of scaling up the magnetic energy on larger scales. It is more important to scale the magnetic field to distribute the magnetic energy according to the $k^{3/2}$ power spectrum, even if the highest energies on the smallest scales are lost due to numerical effects. We note that after the loss of energy, the field strength agrees with that of \cite{Federrath2011} at the corresponding density.

This study does not incorporate the effects of non-ideal MHD. Ohmic, Ambipolar and Hall diffusion are not considered, but contribute to the breakdown of flux freezing. Ohmic dissipation can redistribute magnetic flux from dense regions and enable formation of rotationally supported discs around protostellar cores (e.g. \citealt{Tomida2012}). Ambipolar diffusion causes magnetic decoupling \citep{Desch2001} and creates a magnetic diffusion barrier  in the vicinity of the core, limiting the field amplification and hindering magnetic breaking \citep{Masson2016}. The Hall effect occurs at intermediate densities when neutral collisions preferentially decouple ions from the magnetic field, leaving only the electrons to drift with the magnetic field (e.g. \citealt{ Pandey2008}). However, inclusion of these effects leads to decreased magnetic field strength (e.g. \citealt{Masson2016,Wurster2021}), so the inclusion would not aid in the suppression of fragmentation.

\section{Conclusions}
\label{sec:conclusion}
We have investigated the ability of saturated primordial magnetic fields to suppress the fragmentation of gas during Pop III star formation. We present two-stage MHD zoom-in simulations, whereby random $k^{3/2}$ magnetic fields were superimposed onto the system when the central density achieved 10$^{-13}$ g cm$^{-3}$. We have tracked the fragmentation behaviour of the systems for $\sim 1000$ yr after the formation of the first sink particle. The simulations were repeated for sink particle creation densities in the range 10$^{-10}$-10$^{-8}$ g cm$^{-3}$. Within this range the magnetic pressure remained sub-dominant or comparable with the thermal pressure, providing inadequate support to prevent Jeans instabilities from fragmenting the system. The total number of sink particles formed did not reduce in the magnetised case compared to the purely hydrodynamic scenario. The total mass accreted onto sink particles and the resulting IMFs were also unaffected by the field. As we have not resolved up to stellar core densities, it is possible that the magnetic pressure will become the dominant pressure term in the density range of 10$^{-8}$-10$^{-4}$ g cm$^{-3}$ before the adiabatic core forms, which could suppress fragmentation on the smallest scales not explored in this work. Our results suggest that the inclusion of magnetic fields in numerical simulations of Pop III star formation is unimportant, especially in studies where the maximum resolution is too poor to resolve the scales at which the magnetic pressure could become the dominant support term. Additionally, placing a uniform field over the central collapsing region did result in reduced fragmentation, demonstrating the danger of under-resolving the initial $k^{3/2}$ field in future studies.

\section*{Acknowledgements}
This work used the DiRAC@Durham facility managed by the Institute for Computational Cosmology on behalf of the STFC DiRAC HPC Facility (www.dirac.ac.uk). The equipment was funded by BEIS capital funding via STFC capital grants ST/P002293/1, ST/R002371/1 and ST/S002502/1, Durham University and STFC operations grant ST/R000832/1. DiRAC is part of the National e-Infrastructure.

The authors gratefully acknowledge the Gauss Centre for Supercomputing e.V. (www.gauss-centre.eu) for supporting this project by providing computing time on the GCS Supercomputer SuperMUC at Leibniz Supercomputing Centre (www.lrz.de).

We also acknowledge the support of the Supercomputing Wales project, which is part-funded by the European Regional Development Fund (ERDF) via Welsh Government.

RSK and SCOG acknowledge support from the Deutsche Forschungsgemeinschaft (DFG, German Research Foundation) via the collaborative research center (SFB 881, Project-ID 138713538) ``The Milky Way System'' (subprojects A1, B1, B2 and B8). They also acknowledge support from the Heidelberg Cluster of Excellence ``STRUCTURES'' in the framework of Germany’s Excellence Strategy (grant EXC-2181/1, Project-ID 390900948) and from the European Research Council (ERC) via the ERC Synergy Grant ``ECOGAL'' (grant 855130).

\section*{Data Availability}

The data underlying this article will be shared on reasonable request to the corresponding author.



\bibliographystyle{mnras}
\bibliography{references} 





\bsp	
\label{lastpage}
\end{document}